\documentclass[sigconf,anonymous=false]{acmart}
\pdfoutput=1
\usepackage[latin9]{inputenc}
\usepackage{graphicx}

\makeatletter

\providecommand{\tabularnewline}{\\}

\usepackage{fontawesome}
\usepackage{tikz}
\usetikzlibrary{decorations.text}
\usetikzlibrary{calc,fit}
\usetikzlibrary{arrows,positioning,shapes,snakes,patterns,automata,backgrounds,petri,mindmap,trees,shadows,decorations.shapes}

\usepackage[shortcuts]{extdash}
\usepackage{xspace}
\newcommand{\simt}{\textsc{Sim\==T}\xspace}
\newcommand{\simo}{\textsc{Sim\==O}\xspace}

\copyrightyear{2018}
\acmYear{2018}
\setcopyright{acmcopyright}
\acmConference[NOSSDAV'18]{28th ACM SIGMM Workshop on Network and
Operating Systems Support for Digital Audio and Video}{June 12--15,
2018}{Amsterdam, Netherlands}
\acmPrice{15.00}
\acmDOI{10.1145/3210445.3210453}
\acmISBN{978-1-4503-5772-2/18/06}

\acmArticle{4}

\begin{document}

\author{Pablo Gil Pereira}
\affiliation{%
  \institution{Saarland Informatics Campus}
  \city{Saarbrücken}
  \postcode{66123}
}
\email{gilpereira@cs.uni-saarland.de}

\author{Andreas Schmidt}
\affiliation{%
  \institution{Saarland Informatics Campus}
  \city{Saarbrücken}
  \postcode{66123}
}
\email{andreas.schmidt@cs.uni-saarland.de}

\author{Thorsten Herfet}
\affiliation{%
  \institution{Saarland Informatics Campus}
  \city{Saarbrücken}
  \postcode{66123}
}
\email{herfet@cs.uni-saarland.de}

\renewcommand{\shortauthors}{P. Gil Pereira et al.}

\begin{abstract}
Nowadays Dynamic Adaptive Streaming over HTTP~(DASH) is the most prevalent solution on the Internet for multimedia streaming and responsible for the majority of global traffic.
DASH uses adaptive bit rate~(ABR) algorithms, which select the video quality considering performance metrics such as throughput and playout buffer level.
Pensieve is a system that allows to train ABR algorithms using reinforcement learning within a simulated network environment and is outperforming existing approaches in terms of achieved performance.  
In this paper, we demonstrate that the performance of the trained ABR algorithms depends on the implementation of the simulated environment used to train the neural network. 
We also show that the used congestion control algorithm impacts the algorithms' performance due to cross-layer effects.
\end{abstract}



\makeatother

\title{Cross-Layer Effects on Training \\
Neural Algorithms for Video Streaming}
\maketitle

\section{Introduction}

Considering its share of global bandwidth usage, video streaming is
the most important Internet application nowadays, and is predicted
to increase even more in the next years~\cite{Cisco2017}. These
\emph{over-the-top}~(OTT) services employ \emph{adaptive bit rate}~(ABR)
algorithms to adapt the video download to the current network state,
considering parameters such as the estimated throughput or buffer
occupancy level. When selecting the next bit rate, the goal is to
maximize the user's \emph{Quality of Experience}~(QoE) which depends
on features such as startup delay, number of rebuffering events and
visual quality that is dependent on the used bit rate and bit rate
changes. Until recently, these ABR algorithms were designed with a
specific QoE metric in mind and hence were not usable across different
use cases, for instance on-demand streaming and low-latency applications. 

Pensieve~\cite{Mao2017} is the first solution training an ABR algorithm
using \emph{Reinforcement Learning}~(RL). These ABR algorithms can
adapt to different applications and user preferences by modifying
the reward function that incorporates a QoE metric. While optimizing
the model for perceived performance, Pensieve can learn an optimal
ABR algorithm for different network conditions.

As an application layer protocol, DASH performance is not only dependent
on the ABR algorithm, but also on its interaction with the lower layers,
in particular HTTP and TCP. These cross-layer effects were first discovered
by \cite{Esteban2012,Huang2012}. While the usage of HTTP has an influence
on the performance, this is out of scope for this paper. Instead,
the cross-layer analysis focuses on the TCP protocol, as TCP's \emph{Congestion
Control Algorithms }(CCAs)~\cite{rfc5681} determine the portion
of available throughput the application can use. The cross-layer information
has also been used lately to model QoE~\cite{Stohr2016} and design
ABR algorithms~\cite{Wang2016}. Given that CCAs provide different
performance to DASH applications, Pensieve can consider the cross-layer
effects and learn a different ABR algorithm for each of them.

The contribution of this paper is twofold: First, we extend Pensieve's
network simulator to faithfully handle propagation round-trip time.
Second, we evaluate the cross-layer dependencies between channel characteristics,
TCP congestion control algorithms, QoE metrics, and the resulting
trained ABR algorithms.

The rest of the paper is structured as follows. In Sec.~\ref{sec:Adaptive-Video-Streaming},
we describe how ABR algorithms are learned by Pensieve and the CCAs
that are used for the evaluation. Modifications we applied to Pensieve
are presented in Sec.~\ref{sec:Improving-the-Training} and evaluated
in Sec.~\ref{sec:Evaluation}. We compare our work to other efforts
in the area of cross-layer analysis of DASH performance in Sec.~\ref{sec:Related-Work}.
Finally, we give directions for future research in Sec.~\ref{sec:Future-Work}
and conclude the paper in Sec.~\ref{sec:Conclusions}.

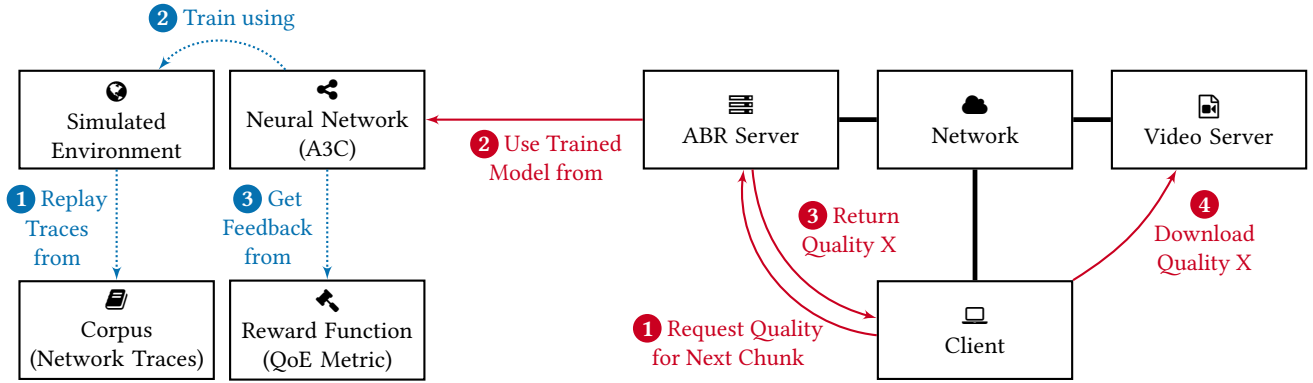
\begin{figure*}
\tikzstyle{b} = [rectangle, draw, fill=white!20, node distance=2.8cm, text width=7.5em, text centered, minimum height=1.3cm, thick]

\definecolor{training_color}{rgb} {0.02,0.443,0.69}
\tikzstyle{t} = [training_color, densely dotted, draw=training_color, text=training_color, -latex',thick]

\definecolor{exection_color}{rgb} {0.792,0.00,0.125}
\tikzstyle{e} = [exection_color, draw=exection_color, text=exection_color, -latex',thick]

\tikzstyle{c} = [draw, thick,line width=0.2em]

\newcommand*\circled[1]{\tikz[baseline=(char.base)]{
            \node[shape=circle,fill,solid,text=white,draw,thick,inner sep=1pt] (char) {\textbf{#1}};}}

\begin{tikzpicture}

    \node [b] (simulator) {\faGlobe\\ Simulated Environment};

    \node [b, right of=simulator] (pensieve) {\faShareAlt\\ Neural Network\\(A3C)};

    \node [b, below of=simulator] (corpus) {\faBook\\ Corpus\\ (Network Traces)};

    \node [b, right of=corpus] (reward) {\faGavel\\ Reward Function\\ (QoE Metric)};

    \node [b, right=6.0cm of pensieve] (network) {\faCloud\\ Network};

    \node [b, left=0.5cm of network] (rl-server) {\faServer\\ ABR Server};

    \node [b, right=0.5cm of network] (video-server) {\faFileVideoO\\ Video Server};

    \node [b, below of=network] (client) {\faLaptop\\ Client};

    \draw [t] (simulator) -- (corpus) node [xshift=-2.5em, midway, align=center] (TextNode) {\circled{1} Replay \\ Traces \\ from };

    \draw [t] (pensieve) to [bend right=50] node [midway, above, align=center] (TextNode) {\circled{2} Train using}  (simulator);

    \draw [t] (pensieve) -- (reward) node [xshift=-2.5em, midway, align=center] (TextNode) {\circled{3} Get \\ Feedback \\ from };

    \draw [e] (client) to [bend left=45] node [midway, xshift=-2.0em, yshift=-2.5em, align=center] (TextNode) {\circled{1} Request Quality\\ for Next Chunk } (rl-server);

    \draw [e] (rl-server) to [] node [midway, xshift=0.5em, yshift=-1.5em, align=center] (TextNode) {\circled{2} Use Trained\\ Model from } (pensieve);

    \draw [e] (rl-server) to [bend right=35] node [midway, xshift=2.5em, yshift=1.5em, align=center] (TextNode) {\circled{3} Return \\ Quality X} (client);

    \draw [e] (client) to [bend right=15] node [midway, xshift=3em, align=center] (TextNode) {\circled{4}\\ Download\\ Quality X } (video-server);

    \draw [c] (client) -- (network);
    \draw [c] (rl-server) -- (network);
    \draw [c] (video-server) -- (network);

\end{tikzpicture}\caption{Pensieve's process is separated into the Training (blue dotted) and
Execution (red solid) phases, which share the trained model for adaptive
streaming.}
\label{fig:pensieve}
\end{figure*}

\section{Adaptive Video Streaming Systems}
\label{sec:Adaptive-Video-Streaming}

From a high level, the performance of a DASH system is measured based
on the quality of the video that is received. This is influenced by
the ABR algorithm, chosing an appropriate quality to request and taking
the buffer occupancy as well as the throughput into account. The latter
is determined by TCP's congestion control, which itself runs a control
loop to probe for available bandwidth and get a fair share. Consequently,
there are two layers with control loops that are usually independent
by design, but entangled at runtime~\cite{Huang2012}.

\subsection{ABR Algorithms and Pensieve }
\label{subsec:ABR-and-Pensieve}

Research in the area of DASH has yielded many different ABR algorithms~\cite{Spiteri2016,Yin2015,shuai2017stabilizing,Wang2016}
as well as QoE metrics~\cite{Liu2015,Alberti2013}. Each ABR algorithm
is designed to maximize the value of a specific QoE metric, so that
in principle we have a direct relation between ABR algorithm and QoE
metric. As soon as a new QoE metric is defined or an existing one
is modified, the ABR algorithm must be updated, which might require
major changes in the bit rate selection strategy.

Pensieve~\cite{Mao2017} is the first solution that trains an ABR
algorithm with network data and uses a given QoE metric to guide the
training. The process is depicted in the left part of Fig.~\ref{fig:pensieve}
and uses reinforcement learning to train with a specific data set,
i.e. a single video with different bit rates. The learning is done
by an A3C network~\cite{Mnih2016}, which uses asynchronous gradient
descent to train neural networks and provides a policy vector as output.
The training input includes information on throughput measurements,
current buffer level, and more parameters. The training uses a corpus
of network traces that are composed of throughput samples over time,
together with a simulator that faithfully models a network with the
respective throughput characteristics. In Sec.~\ref{subsec:propagation_delays},
these traces are extended by a further dimension to increase the fidelity
of the simulator. The rewards are provided to the algorithm by evaluating
a QoE metric, hence measuring the performance of taking a given action
and guiding the algorithm to maximize the value.

After the training, the neural network is used inside the ABR server
depicted in Fig.~\ref{fig:pensieve}. Before requesting a chunk,
the client queries the ABR server and receives the next quality to
be downloaded. Although they can run on different systems for scalability
reasons, we run the client and the ABR server on the same system.

In the following, we use Pensieve in its original as well as a modified
version to learn ABR algorithms and afterwards evaluate them using
the Pensieve DASH client and ABR server.

\subsection{Congestion Control Algorithms}

Over the last decades, many different CCAs have been developed with
the goal of improving stability, fairness, and utilization of TCP
connections over the Internet. The most recent standardized version
is in RFC 5681~\cite{rfc5681} and for instance Linux is using CUBIC~\cite{Ha2008}.
These solutions consider a loss due to retransmission timout as a
signal for congestion and reduce their throughput.

BBR~\cite{Cardwell2016} instead uses measurements of the round-trip
propagation delay~(\emph{$RTprop$}) and the bottleneck bandwidth~($BtlBw$)
to operate at the bandwidth-delay product~($BDP$), which ensures
that the throughput is maximized while the delay is minimized. As
congestion causes queuing delays that result in an increase of RTT
compared to \emph{$RTprop$}, this is used as a signal that congestion
is happening and causes BBR to reduce its sending rate. 

In our evaluation, we run different experiments that use either CUBIC
or BBR and show which effects the selection of a specific CCA can
have on the overall DASH performance in terms of QoE.

\section{Evolving Pensieve}
\label{sec:Improving-the-Training}

Pensieve outperforms existing ABR algorithms regarding the QoE achieved
in the wild~\cite{Mao2017}. We found that this performance can be
improved by modifying the network simulator and extending the network
corpus used for the training.

\subsection{Propagation Delays in the Simulator}
\label{subsec:propagation_delays}

While Pensieve uses traces with samples of throughput over time, it
does not incorporate the round-trip time and in particular the propagation
delay that is independent of the throughput. One of the major differences
between CUBIC and BBR is how they manage their congestion window,
which has an impact on the portion of the RTT that is caused by the
queuing delay.

\subsubsection{Timed Traces}
\label{subsec:Timed-Traces}

Following the data-driven learning approach, it is thus straightforward
to extend the traces in a backwards-compatible way by adding a column
that captures the round-trip propagation time. Consequently, for traces
$i=0,...,n-1$, we store:
\begin{itemize}
\item $t_{i},[t_{i}]=second$: Relative time to the start of the trace.
\item $bw_{i},[bw_{i}]=\frac{Mbit}{second}$: Throughput measured at $t_{i}$.
\item $RTprop_{i},[RTprop_{i}]=second$: Round-trip propagation time measured
at $t_{i}$.
\end{itemize}

\subsubsection{Pensieve's Original Simulator (\simo)}

The original simulator in Pensieve, referred to as \simo in the following,
uses a fixed round-trip propagation delay of 80ms, which is added
to the transmission delay of a video chunk. This transmission delay
depends on the throughput samples in the trace. After the propagation
and transmission delay are summed up, \simo adds $\pm10\%$ to the
total delay, causing uniformly distributed noise. Consequently, the
randomness the simulation faces is proportional to the propagation
delay as well as the transmission delay, so that the transmission
times are also noisy. While propagation delay can vary in the wild,
e.g. as a network card does not have a steady data rate, this additional
noise is high. DASH segments of length $4s$ and encoded at $4Mbps$
take $1s$ at $16Mbps$ line rate of a residential connection. Adding
10\% of noise would result in more than doubling the original delay
of $80ms$. As the variations in line-rate are already part of the
throughput traces, this approach adds this variation twice.

\subsubsection{Modified Simulator with Timed Traces (\simt)}
\label{subsec:sim-t}

In order to investigate the impact of propagation delay handling within
the simulator, we first augmented the traces, which resulted in a
structure as in Sec.~\ref{subsec:Timed-Traces}. We used the corpus
described in Sec.~\ref{subsec:Pensieve-corpus-RTT} which uses the
same base $RTprop$ as the original simulator.

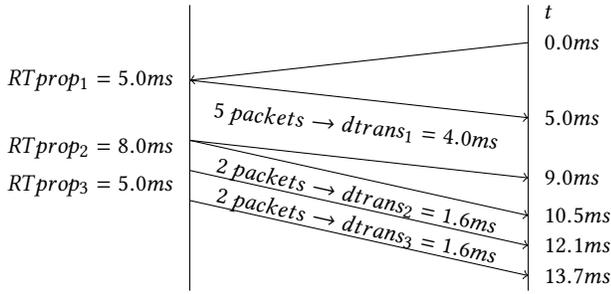
\begin{figure}
\begin{tikzpicture}
[my shape/.style={rectangle split, rectangle split parts=#1}]
\tikzset{input/.style={}}

\draw (0,0)--++(0,-3.8cm);
\draw (4.5,0)--++(0,-3.8cm);

\draw[->] (0,-1.0) coordinate (a) --
node [near end, xshift=-3em, sloped, below=0,my shape=2, minimum height=0.75cm, rectangle split horizontal] (dt1) {$5\;packets \rightarrow dtrans_1=4.0ms$}
++(4.5,-0.5) coordinate (b);


\node [left =1mm of a,my shape=1, rectangle split horizontal] (dp1) {$RTprop_1=5.0ms$};

\node [left =1mm of a,yshift=-0.9cm,my shape=1, rectangle split horizontal] (dp2) {$RTprop_2=8.0ms$};

\node [left =1mm of a,yshift=-1.4cm,my shape=1, rectangle split horizontal] (dp3) {$RTprop_3=5.0ms$};

\node [right=1mm of b, my shape=1, yshift=1.4cm, rectangle split horizontal] (tt)  {$t$};

\node [right=1mm of b, my shape=1, yshift=1.0cm, rectangle split horizontal] (t0)  {$0.0ms$};

\node [right=1mm of b, my shape=1, yshift=0.0cm, rectangle split horizontal] (t1)  {$5.0ms$};

\node [right=1mm of b, my shape=1, yshift=-0.8cm, rectangle split horizontal] (t2)  {$9.0ms$};

\node [right=1mm of b, my shape=1, yshift=-1.3cm, rectangle split horizontal] (t3)  {$10.5ms$};

\node [right=1mm of b, my shape=1, yshift=-1.7cm, rectangle split horizontal] (t4)  {$12.1ms$};

\node [right=1mm of b, my shape=1, yshift=-2.1cm, rectangle split horizontal] (t5)  {$13.7ms$};

\draw[->] (4.5,-0.5) coordinate (a) -- ++(-4.5,-0.5) coordinate (b);

\draw[->] (0,-1.8) coordinate (a) -- ++(4.5,-0.5) coordinate (b);

\draw[->] (0,-1.8) coordinate (a) --
node [near end, xshift=-3em,sloped, below=0mm,my shape=2, minimum height=0.40cm, rectangle split horizontal] (dt2) {$2\;packets \rightarrow dtrans_2=1.6ms$}
++(4.5,-1.0) coordinate (b);

\draw[->] (0,-2.2) coordinate (a) --
node [near end, xshift=-3em,sloped, below=0mm,my shape=2, rectangle split horizontal] (dt3) {$2\;packets \rightarrow dtrans_3=1.6ms$}
++(4.5,-1.0) coordinate (b);

\draw[->] (0,-2.6) coordinate (a) -- ++(4.5,-1.0) coordinate (b);

\end{tikzpicture}

\caption{Model of overall delays with varying propagation delays for packets
of 1500 bytes sent at 15Mbps.}
\label{fig:delays}
\end{figure}

The $RTprop$ a complete chunk faces is computed as the propagation
time for the request and the maximum of propagation delays faced by
the responses. This is further motivated by the exemplary transmission
in Fig.~\ref{fig:delays}, where an initial $RTprop$ of 5ms increases
to 8ms and returns back to 5ms after a handful of packets. The transmission
delay of the chunk is the sum of transmission delays for the individual
packets, which vary with the throughput from the trace. Hence, we
get the following delays for chunk $i$ which was transmitted using
packets $p=0,...,n-1$, each of size P:

\begin{equation}
RTprop(i)=\frac{RTprop_{0}}{2}+\max_{p=0,...,n-1}(\frac{RTprop_{p}}{2})\label{eq:max-filter}
\end{equation}

\begin{equation}
dtrans(i)=\sum_{p=0}^{n-1}\frac{P}{bw_{p}}
\end{equation}

While this model does not consider reordering of packets in a flow,
e.g. by route changes, it models the actual behavior of a channel
more closely and still keeps the simulation straightforward. Assuming
the client selects a bit rate roughly the same as the fair share,
a chunk download takes the length of the chunk in seconds, which is
of an order of magnitude smaller than the route change period, so
that we can assume the route stays constant. 

\subsection{Network Trace Corpora}

We have used different corpora of network traces for the evaluations
in Sec.~\ref{sec:Evaluation}. Firstly, the Pensieve corpus \cite{Mao2017}
and variations of it, which were created using publicly available
datasets. Secondly, what we call the DASH corpora, as they have been
generated using traces from DASH-IF clients running on the testbed
introduced in Sec.~\ref{subsec:Scenarios}. Each corpus is separated
into training and test sets to avoid overfitting and evaluate the
performance of resulting algorithms on data they have not processed
before.

\subsubsection{Pensieve Corpus}
\label{subsec:Pensieve-Corpus}

This is the corpus used in \cite{Mao2017}, in the following referred
to as $C_{P}$. Two public datasets were used to compose it with broadband
and mobile Internet traffic, respectively. The traces are taken from
the ``web browsing'' category in both datasets, meaning only HTTP
downloads are considered, which range from regular, static web pages
to streamed video.

\subsubsection{Pensieve with RTprop Corpus}
\label{subsec:Pensieve-corpus-RTT}

In order to incorporate the propagation delay, we have augmented the
original traces with a base $RTprop$ of $80ms$ and added $\pm10\%$
uniformly distributed random noise, which yields the corpus $C_{RTprop}$.
Considering the original implementation of the simulator, this is
the corpus to be used to mimic the original behavior of Pensieve as
closely as possible.

\subsubsection{DASH Corpora}
\label{subsec:DASH-Corpora}

Several CCAs are currently available, which achieve different performance
with DASH, as shown in \cite{Stohr2016} and analyzed in Sec.~\ref{subsec:cross-layer-pensieve}.
In order to evaluate whether Pensieve can circumvent cross-layer effects,
we ensure that only the performance achieved by a single CCA is represented
in the traces. Two DASH corpora are generated complying with these
requirements, namely \emph{$C_{CUBIC}$} and $C_{BBR}$, which are
generated only considering traces collected with CUBIC and BBR, respectively.

These traces are collected in a controlled environment, which is described
in Sec.~\ref{subsec:Scenarios}. In order to avoid overfitting for
a single throughput, we have configured the link with \{3,~3.5,~4,~4.5\}~Mbps,
while the latency is always $80ms$.\textbf{ }These corpora are generated
just considering throughput and latency samples achieved by the DASH
client, as DASH has an on-off traffic pattern whose performance differs
from other applications~\cite{Huang2012}. The traces were collected
25 times for periods of 5 minutes. 

We assume there is a correlation between consecutive throughput or
latency traces within a download. For instance, if the congestion
window is increased beyond the channel's BDP, that would result in
a higher share of the available throughput, but this results in a
higher RTT because of a queueing delay at the bottleneck buffer. Therefore,
the traces in these corpora are not randomly selected, because this
removes the correlation information and results in pathological, unrealistic
network behaviours. In order to keep this information, the traces
are collected in the sequence they were originally captured within
the video download, with 1s granularity.

Despite not being collected in a realistic environment, the DASH corpora
fulfill their purpose of representing the performance DASH applications
achieve depending on the underlaying congestion control. Given that
BBR and CUBIC also achieve different performance in the wild~\cite{Cardwell2016},
we think the results in Sec.~\ref{subsec:cross-layer-pensieve} also
hold for traces collected in a more realistic scenario. In Sec.~\ref{sec:Future-Work}
we mention methods to collect realistic traces sorted by CCA.

\subsection{Entropy Weights and Training Duration}

Pensieve uses the gradient of the entropy of the policy vector as
regularization term to avoid overfitting. The weight of the regularization
term is a major hyperparameter for the learning phase, as the model
tends to prematurely converge to suboptimal policies at the beginning
of the training. The authors in~\cite{Mao2017} suggest starting
with a high entropy factor of approx. 1.0 and progressively reduce
it to 0.1 over 50,000 iterations. However, they do not specify how
the entropy factor is changed over time. Having tested stepwise and
linear approaches , we get the best performance when the models are
trained with linear decrease from 1.0 to 0.1 with steps of 0.01 over
100,000 iterations. Thus, all the models presented in this paper have
been trained with this policy. The selection of the optimal number
of iterations and entropy weight progression is out of scope for this
paper, but we are confident that finding such a scheme can also improve
what our models can achieve.

\section{Evaluation}
\label{sec:Evaluation}

With the changes applied to Pensieve, we evaluate the resulting performance
and consider the cross-layer effects that the choice of a given QoE
metric and CCA have on the overall performance.

\subsection{Methodology}
\label{subsec:Methodology}

For the experiments, we compare different QoE metrics, use different
network scenarios and characteristics, and employ the dataset from
the Pensieve paper.

\subsubsection{QoE Metrics}
\label{subsec:QoE-Metrics}

In order to measure the performance of the neural network we use the
general QoE metric defined in Eq.~\ref{eq:general-qoe} for our experiments~\cite{Yin2015}.
For a total number of $N$ chunks, the metric considers each chunk's
bitrate $R_{n}$ and rebuffering time $T_{n}$, which results from
the download of that chunk. The function $q(R_{n})$ is the bitrate
utility that maps the bitrate $R_{n}$ to the quality perceived by
the user. The last term penalizes video quality switches to favor
smoothness. Finally, $\mu$ is the rebuffering penalization term. 

\begin{equation}
QoE=\sum_{n=1}^{^{N}}q(R_{n})-\mu\sum_{n=1}^{^{N}}T_{n}-\sum_{n=1}^{^{N-1}}\left|q(R_{n+1})-q(R_{n})\right|\label{eq:general-qoe}
\end{equation}

We have used two different bit rate utility functions, resulting in
the following two QoE metrics:
\begin{itemize}
\item $QoE_{lin}$: Linear mapping, where the utility function is the chunk's
bit rate. Therefore, we set $q(R_{n})=R_{n}$ and $\mu=4.3$ as in
\cite{Yin2015}.
\item $QoE_{HD}$: This metric assigns higher values to \emph{high definition}~(HD)
video than it does for lower qualities. The bit~rate-to-quality mapping
for the Pensieve dataset can be found in Tab.~\ref{tab:bitrate-quality}
and sets $\mu=8$ as in \cite{Mao2017}.
\end{itemize}
\begin{table}
\begin{tabular}{|c|c|c|c|c|c|c|}
\hline 
$R_{n}(Mbps)$ & 0.3 & 0.75 & 1.2 & 1.85 & 2.85 & 4.3\tabularnewline
\hline 
$q(R_{n})$ & 1 & 2 & 3 & 12 & 15 & 20\tabularnewline
\hline 
\end{tabular}

\caption{Mapping of $\mathbf{QoE_{HD}}$ bit rates to quality levels.}
\label{tab:bitrate-quality}
\end{table}

\subsubsection{Scenario}
\label{subsec:Scenarios}

The testbed uses IPSec and L2TP tunnels through the GEANT\footnote{\url{https://www.geant.org/}}
research network for connecting the two remote locations and OpenvSwitch
to do the bridging. All hosts are VirtualBox virtual machines (VM)
running the same version of Ubuntu and they are connected to the nodes
via VirtualBox host-only interfaces, which are then bridged to OpenvSwitch.
Across the L2TP tunnel we apply network emulation~(\texttt{netem})
to add delays to achieve a total round-trip time of $80ms$, as well
as for packet loss emulation for Sec.~\ref{subsec:pensieve-losses},
and token bucket filter~(\texttt{tbf}) to limit the throughput to
different values.

\subsubsection{Dataset}

For all the experiments we have used the same video dataset used in
~\cite{Mao2017}, which is encoded by the H.264/MPEG-4 codec at the
bitrates \{0.3,~0.75,~1.2,~1.85,~2.85,~4.3\} Mbps. The video,
which has a total lenght of 193 seconds, has been divided into 48
chunks of 4 seconds each and a last chunk of 1 second.

\subsection{Implementation}

The experiment has three separate components, namely the client, the
ABR server and the DASH server. While the DASH server runs in a dedicated
VM, the DASH client and ABR server run on the same VM. The operating
system in all the VMs is Ubuntu 16.04 LTS with Linux kernel 4.9, which
is the first kernel including BBRv1.0~\cite{Cardwell2016}.

The DASH server is an Apache server v2.4.18 with the cross-origin
resource sharing~(CORS) capabilities enabled. The ABR server is presented
in~\cite{Mao2017}, which uses \emph{BaseHTTPServer}\footnote{\url{https://docs.python.org/2/library/basehttpserver.html}}
to handle HTTP requests from the client and \emph{TensorFlow} to execute
the Actor network. Finally, we used the DASH-IF client provided in
the Pensieve repository\footnote{\url{https://github.com/hongzimao/pensieve}},
which requests the next bitrate to the ABR server instead of using
the default offline ABR algorithm \cite{Mao2017}.

\subsection{\simo vs. \simt}
\label{subsec:sim-comparison}

In order to show the impact of propagation delay handling, we compared
the performance of ABR algorithms trained using Pensieve's original
simulator~(\simo) and those trained using our simulator with timed
traces~(\simt), which were presented in Sec.~\ref{subsec:propagation_delays}.
For both simulators we used $QoE_{HD}$, which is an aggressive metric.
For this experiment we use the testbed described in Sec.~\ref{subsec:Scenarios}
with 80ms RTT and two throughput configurations: a limited scenario
with 3Mbps, where the available throughput is smaller than the highest
video bit rate, and an unlimited scenario with 6Mbps. The QoE samples
have been collected for 5 minutes for each execution. 

\begin{figure}
\includegraphics[width=1\columnwidth]{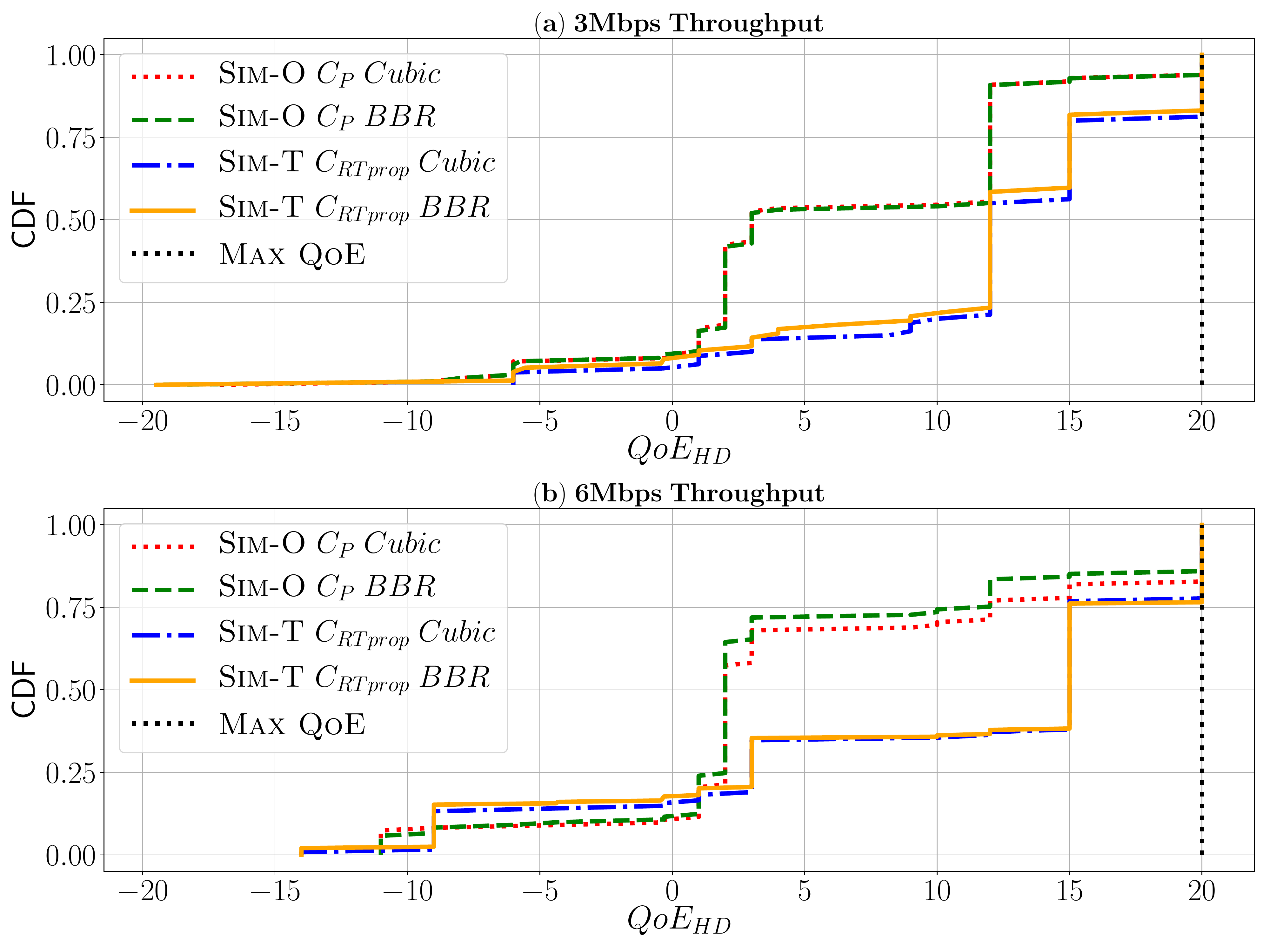}

\caption{Comparison of \simo and \simt with $QoE_{HD}$ and bottleneck bandwidth
of 3Mbps and 6Mbps.}
\label{fig:qoe_hd-1}
\end{figure}

The models resulting from training with $QoE_{HD}$ have a better
performance when trained with \simt instead of \simo (Fig.~\ref{fig:qoe_hd-1}).
Therefore, the proper handling of RTTs in the simulator allows more
aggressive decisions, which in turn results in better user experience.
ABR algorithms learned using \simt are trained in a more faithful
streaming environment, which leads to a situation where it can make
better decisions than those that use \simo. It is important to note
that this performance difference is only due to a change in the simulator
and without any changes in the neural network used to learn the ABR
algorithm.

\subsection{Cross-layer Effects with Pensieve}
\label{subsec:cross-layer-pensieve}

Previous work shows cross-layer effects on DASH depending on the TCP
layer, so we are going to investigate the effects of using different
TCP algorithms on the achieved QoE for learned ABR algorithms. We
train ABR algorithms on corpora that used a single CCA and show that
this has an impact on the resulting performance.

\subsubsection{Channel Loss}
\label{subsec:pensieve-losses}

\begin{figure}
\includegraphics[width=1\columnwidth]{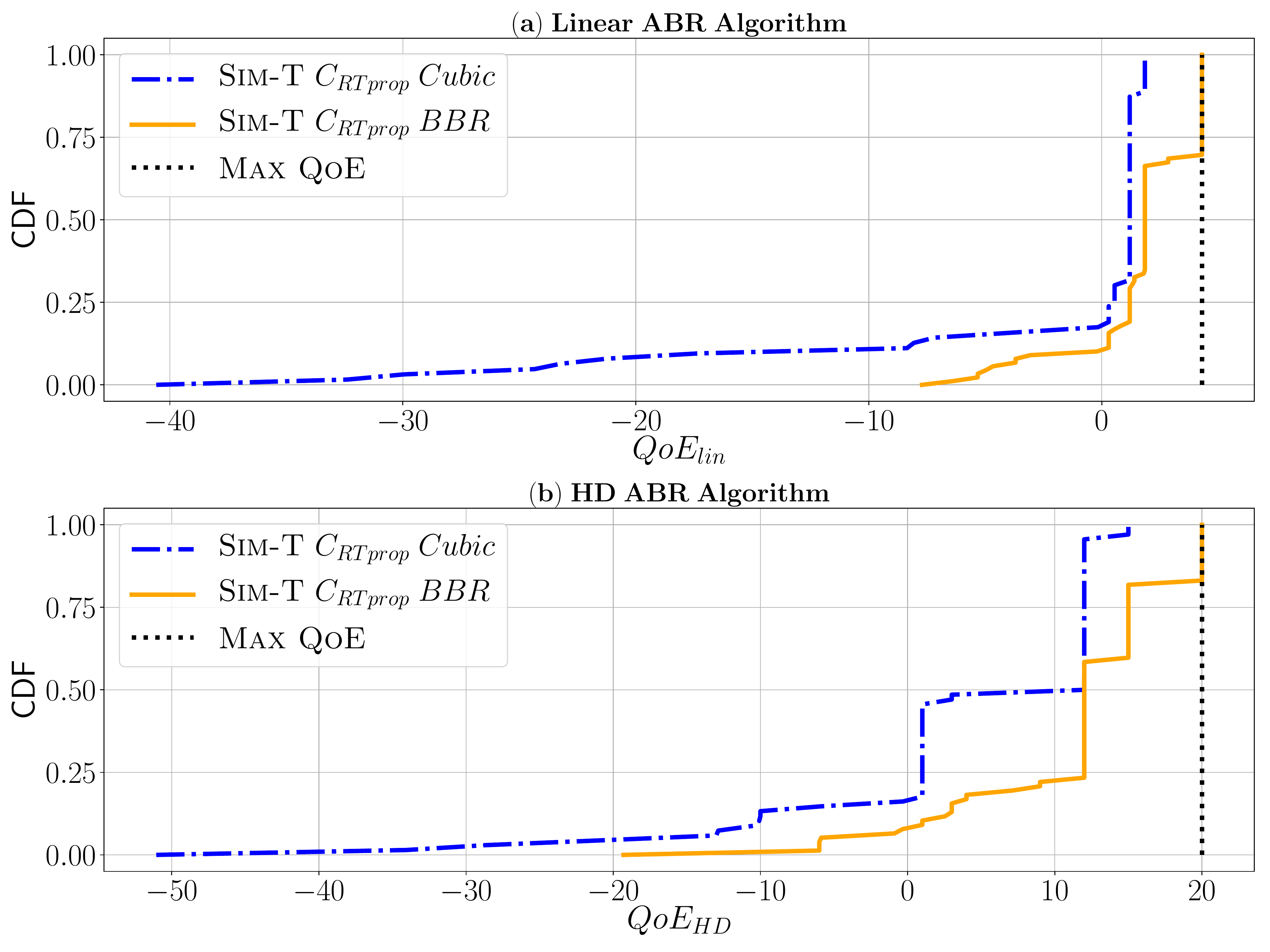}

\caption{Scenarios with 2\% loss lead to significant difference in the performance
of ABR algorithms, depending on the used CCA and QoE metric.}
\label{fig:BBR-Cubic-losses}
\end{figure}

A major difference between BBR and CUBIC is the throughput they can
sustain in scenarios with significant channel loss, as shown in~\cite{Cardwell2016}.
For evaluating its impact on DASH applications, we compare how models
trained with \simt perform with 2\% packet loss. We omit \simo in
this analysis as Sec.~\ref{subsec:sim-comparison} shows that the
training with \simt results in better performance. 

We can see in Fig.~\ref{fig:BBR-Cubic-losses} that DASH achieves
better performance using BBR as the CCA, independently of the used
QoE metric. This is due to the stable throughput BBR can achieve on
lossy channels, as it does not reduce the congestion window with a
packet loss. There is a major gap between CUBIC and BBR with $QoE_{HD}$.
This is in line with the results in~\cite{Huang2012}, which show
that DASH needs to choose bit rates aggressively to allow TCP to effectively
probe for the available throughput. This is the case for $QoE_{HD}$,
which selects higher bit rates leading to higher rewards during training.

\subsubsection{Congestion Control Algorithm}
\label{subsec:cca-algorithms}

\begin{figure}
\includegraphics[width=0.9\columnwidth]{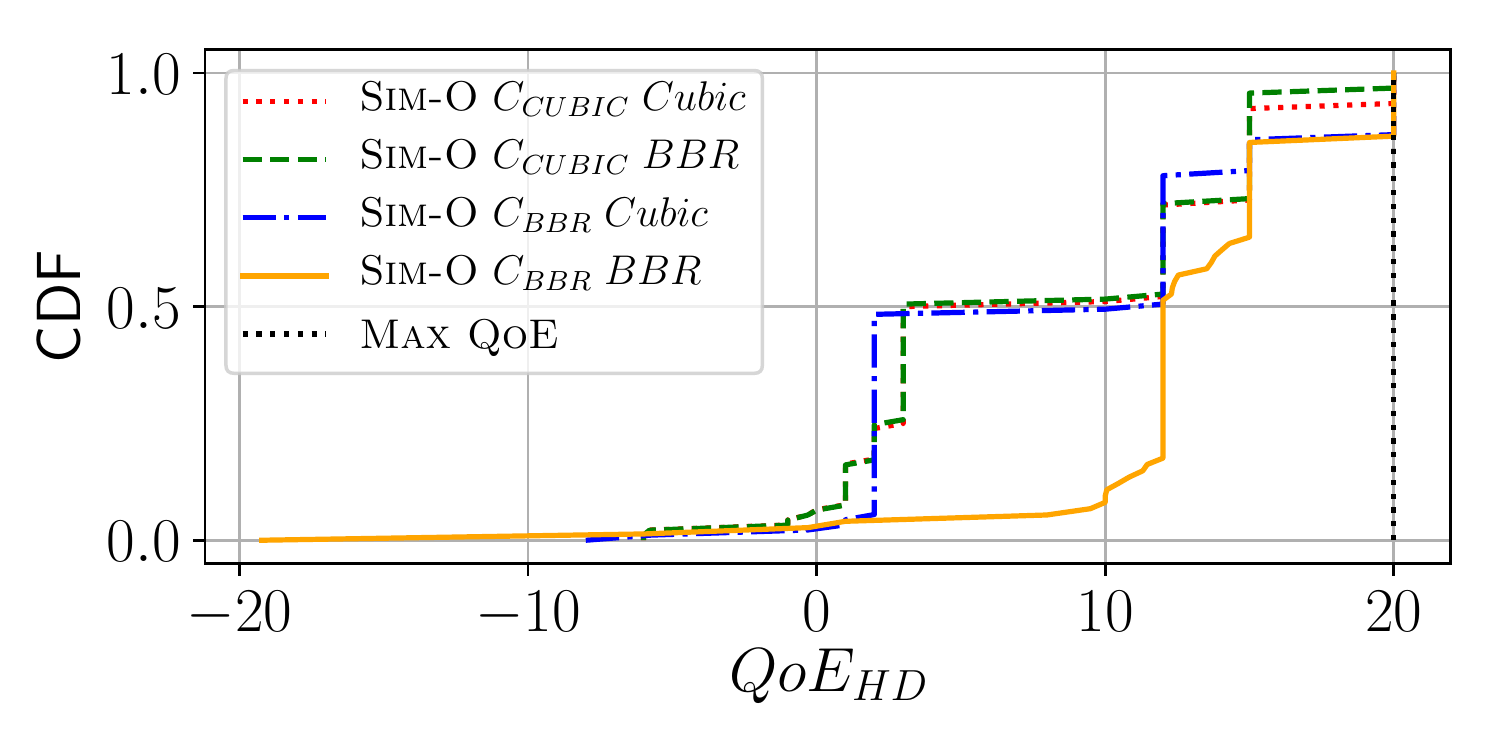}

\caption{Comparison of ABR algorithms trained on CUBIC and BBR DASH corpora
and executed with CUBIC and BBR indicate cross-layer effects
with neural adaptive streaming.}

\label{fig:cubic-vs-bbr-corpora}
\end{figure}

Using the $C_{CUBIC}$ and $C_{BBR}$ corpora, which are described
in Sec.~\ref{subsec:DASH-Corpora}, we trained two instances of \simo
to compare the performance of the resulting algorithms. The scenario
in Sec.~\ref{subsec:Scenarios} is configured with 3Mbps throughput,
80ms RTT, and no loss.

Fig.~\ref{fig:cubic-vs-bbr-corpora} shows that the model with the
best performance is trained with $C_{BBR}$ and executed with BBR.
As BBR aims to operate at the BDP, it can achieve a more constant
throughput, as the congestion window is not reduced with losses caused
by bandwidth probing. \simo with $C_{CUBIC}$ executed with BBR is
not able to exploit BBR's desirable characteristic, as the model learned
to run on top of CUBIC, resulting in the CUBIC-like performance. Fig.~\ref{fig:cubic-vs-bbr-corpora}
also shows that a mismatch of the CCA between traces collection and
client execution might lead to poor performance.

In consequence, DASH performance depends on the underlying CCA as
Pensieve learned a different ABR algorithm for BBR and CUBIC, so cross-layer
effects have to be considered. As a result of their impact in the
performance of the learned ABR, we think CCAs should be considerend
for a more realistic traces collection.

\section{Related Work}
\label{sec:Related-Work}

Cross-layer effects for \emph{HTTP Adaptive Streaming }(HAS), were
first discovered by Esteban et al.~\cite{Esteban2012}, who describe
the effects during the different phases of TCP congestion control,
namely initial burst, ACK clocking and trailing ACK phases. They show
that loss is most damaging in the trailing ACK phase, where packets
are still in flight but no new data can be sent. They also analyze
how pacing avoids burst losses, but report only minimal throughput
gains.

Cross-layers effects also impact the design of ABR algorithms, as
introduced by Huang et al.~\cite{Huang2012}. They point out that
DASH ``on-off'' traffic scheduling prevents TCP for probing for
its fair share. Thus, ABR algorithms should not rely on throughput
estimation to select the bit rate. This is further investigated in
Wang et al.~\cite{Wang2016}, where an ABR algorithm based on cross-layer
effects is presented.

The extensive emulation and analysis of cross-layer dependencies by
Stohr et al.~\cite{Stohr2016} found that there is a linear relation
between delay, loss and QoE, that is independent of the bandwidth
- a result that is further supported by our findings regarding the
incorporation of latency information into the training. As a consequence,
a corpus for Pensieve considering more features than only throughput
is necessary, as well as changes in the neural network's input.

Another analysis of cross-layer behavior by Bhat et al.~\cite{Bhat2017}
investigates how replacing the TCP and HTTP layer within DASH can
change the overall performance of DASH, which was also investigated
by Timmerer et al.~\cite{Timmerer2016}. The authors show that unmodified
DASH clients are not able to improve QoE when compared to DASH over
TCP and even result in lower bit rates.

\section{Future Work}
\label{sec:Future-Work}

While Pensieve can learn ABR algorithms that outperform existing ones,
there are opportunities for future extensions beyond what the original
paper describes.

\textbf{Propagation Delay as Model Input.} We have shown how important
the propagation delay is for proper simulation of a network environment.
We propose to extend the state vector Pensieve uses by measurements
of the current round-trip propagation time. This RTT information becomes
even more important when buffer size and chunk sizes shrink, which
is the case for low-latency applications and hardware that incorporates
a fixed amount of memory. 

\textbf{Leveraging (More) Cross-Layer Information. }As Huang et al.~\cite{Huang2012}
show, throughput should be measured by TCP itself and not within DASH.
The same holds for other parameters such as losses or propagation
delay, which can be measured by TCP implementations and used by the
model as additional state information features. BBR for instance has
a faithful model of the channel that can be used by the ABR algorithm
for a better bit rate selection. 

\textbf{Extended Network Corpus. }We think that actual network measurements
for HTTP traffic that incorporate delay information are required,
as well as traces sorted by CCA as shown in Sec.~\ref{subsec:cca-algorithms}.
Gathering this corpus could for instance be done using the RIPE Atlas\footnote{\url{https://atlas.ripe.net/}}
project that allows to run HTTP measurements from a wide variety of
end-systems to well-defined IP addresses or hostnames.

\textbf{Reinforcement Learning for Low-latency Streaming. }Pensieve,
like most of the ABR algorithms, relies on a large playback buffer
to overcome TCP throughput fluctuations. However, large buffers are
not suitable for live streaming, since the video is not completely
available in advance. In scenarios with such low-latency constraints,
the buffer level becomes more important, as small fluctuations may
result in re-bufferings. We suggest extending the state vector so
that not only the last buffer level sample, but a short history of
it is sent to the ABR server, which would allow to track how the application
reads the buffer.

\section{Conclusion}
\label{sec:Conclusions}

Training ABR algorithms for DASH using neural networks proves to yield
outstanding results. Nevertheless, this performance can be improved
even further by making the simulator used during training more realistic.
As we have shown, replacing a randomized treatment of link round-trip
times with a trace-based approach can help to achieve better results.
We have also shown in our cross-layer analysis how the CCA algorithm
impacts the performance of ABR algorithms. Finally, we predict that
extending the training model by additional input signals and gather
these signal from the transport layer can lead to a further increase
of reliability and performance of neural adaptive streaming systems.

\begin{acks}

{\small{}The work is supported by the \grantsponsor{}{German Research
Foundation (DFG)}{} as part of SPP 1914 ``Cyber-Physical Networking''
under grant \grantnum{}{HE~2584/4-1}.}{\small\par}

\end{acks}

\bibliographystyle{plain}
\bibliography{ms}

\end{document}